\documentclass[pre,twocolumn]{revtex4} %,preprint,endfloats

\usepackage{t1enc,alltt,amssymb,amsmath,natbib}
\usepackage{graphicx}

\begin{document}

\title{Transitions in the wake of a flapping foil}

\author{Ramiro Godoy-Diana}
  \email{ramiro@pmmh.espci.fr}
\author{Jean-Luc Aider}
\author{Jos\'e Eduardo Wesfreid}
 \affiliation{{Physique et M\'ecanique des Milieux H\'et\'erog\`enes}\\
{(PMMH) UMR 7636 CNRS, ESPCI, Paris 6, Paris 7}\\
{10, rue Vauquelin, F-75231 Paris Cedex 5, France}\\}

\begin{abstract}
We study experimentally the vortex streets produced by a flapping foil in a hydrodynamic tunnel, using 2D Particle Image Velocimetry (PIV). An analysis in terms of a flapping frequency-amplitude phase space allows to identify: 1) the transition from the well-known B\'enard-von K\'arm\'an (BvK) wake to the reverse BvK vortex street that characterizes propulsive wakes, and 2) the symmetry breaking of this reverse BvK pattern giving rise to an asymmetric wake. We also show that the transition from a BvK wake to a reverse BvK wake precedes the actual drag-thrust transition and we discuss the significance of the present results in the analysis of flapping systems in nature.
\end{abstract}

\maketitle

\section{Introduction}

Control of vortices produced by flapping wings or fins to generate propulsive forces is the everyday task of birds, insects and swimming animals. Many studies of actual flapping extremities have been driven by the need for a better understanding of this form of propulsion with the ultimate goal of enhancing man-made propulsive devices \cite[see extensive reviews in][]{triantafyllou2000,rozhdestvensky2003,triantafyllou2004,wang2005,fish2006}. A wake vortex street with the sign of vorticity of each vortex reversed with respect to the typical B\'enard-von K\'arm\'an (BvK) vortex street is an ubiquitous feature related to propulsive flapping motion. Such {\em reverse BvK vortex streets} have not only been observed in the wakes of swimming fish \cite[see e.g.][]{wolfgang1999,drucker_lauder2001} but also studied in detail through laboratory experiments with flapping foils \cite[][]{koochesfahani1989,gopalkrishnan1994,anderson1998,jones1998,lai1999,vandenberghe2004,parker2005,buchholz2006} and numerical simulations \cite[][]{jones1998,zhu2002,guglielmini2004,alben2005,blondeaux2005,dong2006}. The mean velocity profile associated with these thrust-producing reverse BvK streets has the form of a jet and the stability properties of this average jet flow seem closely related to the propulsive efficiency of the flapping foil \cite{triantafyllou1991}.

Flapping systems, either natural or man-made, are often discussed in terms of a single parameter, the Strouhal number, defined as the product of the flapping frequency $f$ and amplitude $A$ divided by the cruising speed $U$, i.e. $\mathrm{Sr}_A=fA/U$. Experiments with flapping foils \cite[][]{anderson1998,read2003} have shown that propulsive efficiency peaks within a narrow interval $0.2<\mathrm{Sr}_A<0.4$. Data compiled by \cite{triantafyllou1991,taylor2003} shows that observed cruise Strouhal numbers for a wide range of flying and swimming animals also lie on this interval, suggesting that, not surprisingly, natural selection has tuned animals for high propulsive efficiency. However, less clear are the physical reasons that determine certain flapping configurations, with their associated vortex signature in the wake, to be the optimal choices for efficient thrust generation.

Although three-dimensionality plays an important role in most flapping systems, a quasi-two-dimensional picture of the wake often contains key dynamical elements, such as the creation and organization of vorticity, that are crucial in every system involving flapping foils as a means of producing propulsive forces \cite[][]{wang2000prl,minotti2002,protas2003,alben2005}. The goal of the present work is to explore these quasi-2D mechanisms using a simple but very well controlled experiment, in order to provide a definitive framework for the analysis of the transitions observed in the wake of flapping foils.

\section{Experimental setup and parameters}

\begin{figure}[b!]
\centering
\includegraphics[width=0.7\linewidth]{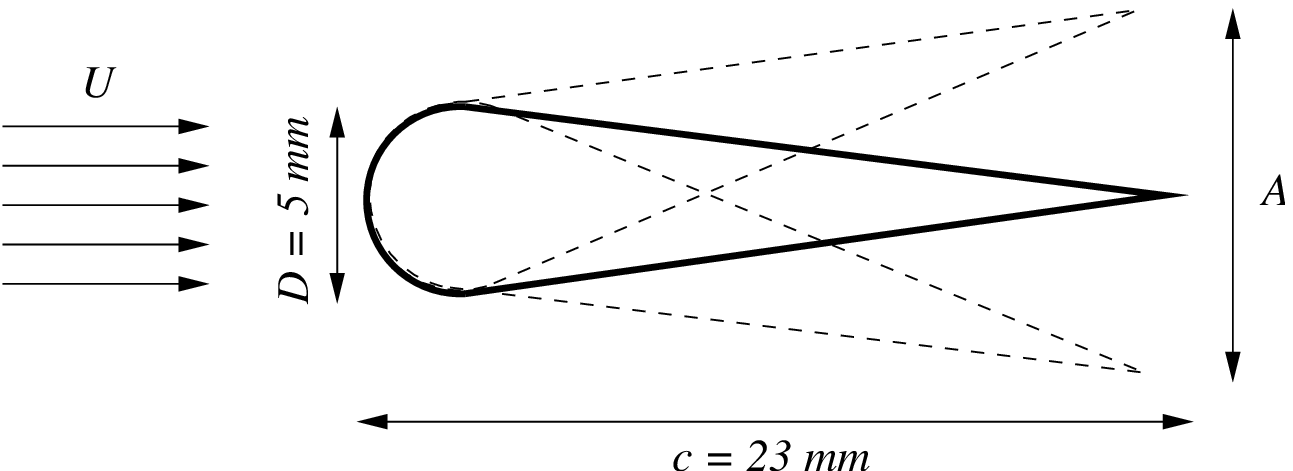}
\includegraphics[width=0.95\linewidth]{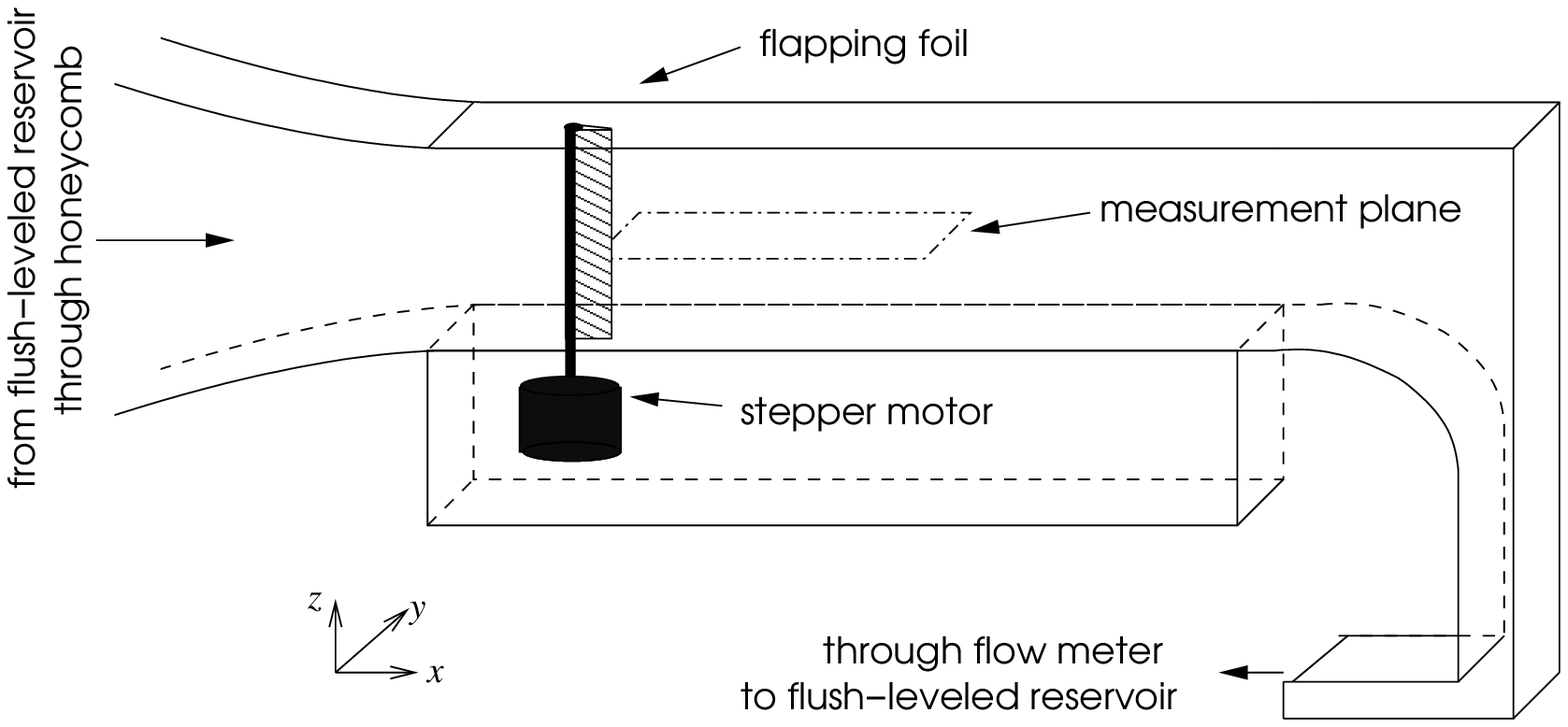}
\caption{Schematic views of the foil profile and the hydrodynamic tunnel. The foil chord $c$ is 23mm and its span is 100mm which covers the whole height of the 100 $\times$ 150mm section of the tunnel. The foil profile is symmetric, opening at the leading edge as a semicircle of diameter $D=5$mm which is also the maximum foil width. The pitching axis is driven by a stepper motor.}
\label{fig_setup}
\end{figure}
The setup consists of a pitching foil of aspect ratio 4 placed in a hydrodynamic tunnel (see Fig. \ref{fig_setup}). The experimental control parameters are the flow velocity in the tunnel $U$, the foil oscillation frequency $f$ and peak-to-peak amplitude $A$. The Reynolds number $\mathrm{Re}=UD/\nu$, where $D$ is the foil width (or thickness) and $\nu$ is the kinematic viscosity, was set to 255 in the experiments reported here. This corresponds to 1173 for the chord-based Reynolds number $\mathrm{Re}_c=Uc/\nu$, which is an intermediate value in the $100 < \mathrm{Re}_c < 10000$ range of `intermediate Reynolds numbers' that characterize flapping-based propulsion in nature \cite{wang2005}. In addition, we define a dimensionless flapping amplitude $A_D$ and a Strouhal number $\mathrm{Sr}$ as

\begin{equation}\label{strouhal}
A_D=A/D \;\;\; \mathrm{and} \;\;\; \mathrm{Sr}=fD/U \; .
\end{equation}

\noindent Although natural vortex shedding exists for the steady flap at this Reynolds number even at zero-angle of incidence \footnote{The critical Reynolds number for the flap is approximately 140.}, no mode competition is observed in the strongly forced flapping regimes studied here. The flapping frequency used to define $\mathrm{Sr}$ is thus equivalent to the main vortex shedding frequency. In addition, we have chosen to use a fixed length scale ($D$) instead of the usual peak-to-peak amplitude ($A$) in the definition of $\mathrm{Sr}$ so that each degree of freedom of the flapping motion is represented in one non-dimensional parameter. The $(\mathrm{Sr}, A_D)$ phase space thus defined allows to give a clear picture of the different observed regimes. Measurements were performed using 2D Particle Image Velocimetry (PIV) on the horizontal mid-plane of the flap. PIV was performed using a LaVision system with an ImagerPro $1600 \times 1200$ 12-bit charge-coupled device (CCD) camera recording pairs of images at $\sim{15}$Hz and a two rod Nd:YAG (15mJ) pulsed laser. Laser sheet width was about 1mm in the whole 100mm $\times$ 80mm imaging region. The time lapse between the two frames (d$t$) was set to 12ms.

\begin{figure*}
\centering
\includegraphics[width=0.85\linewidth,viewport=77 86 531 772,clip]{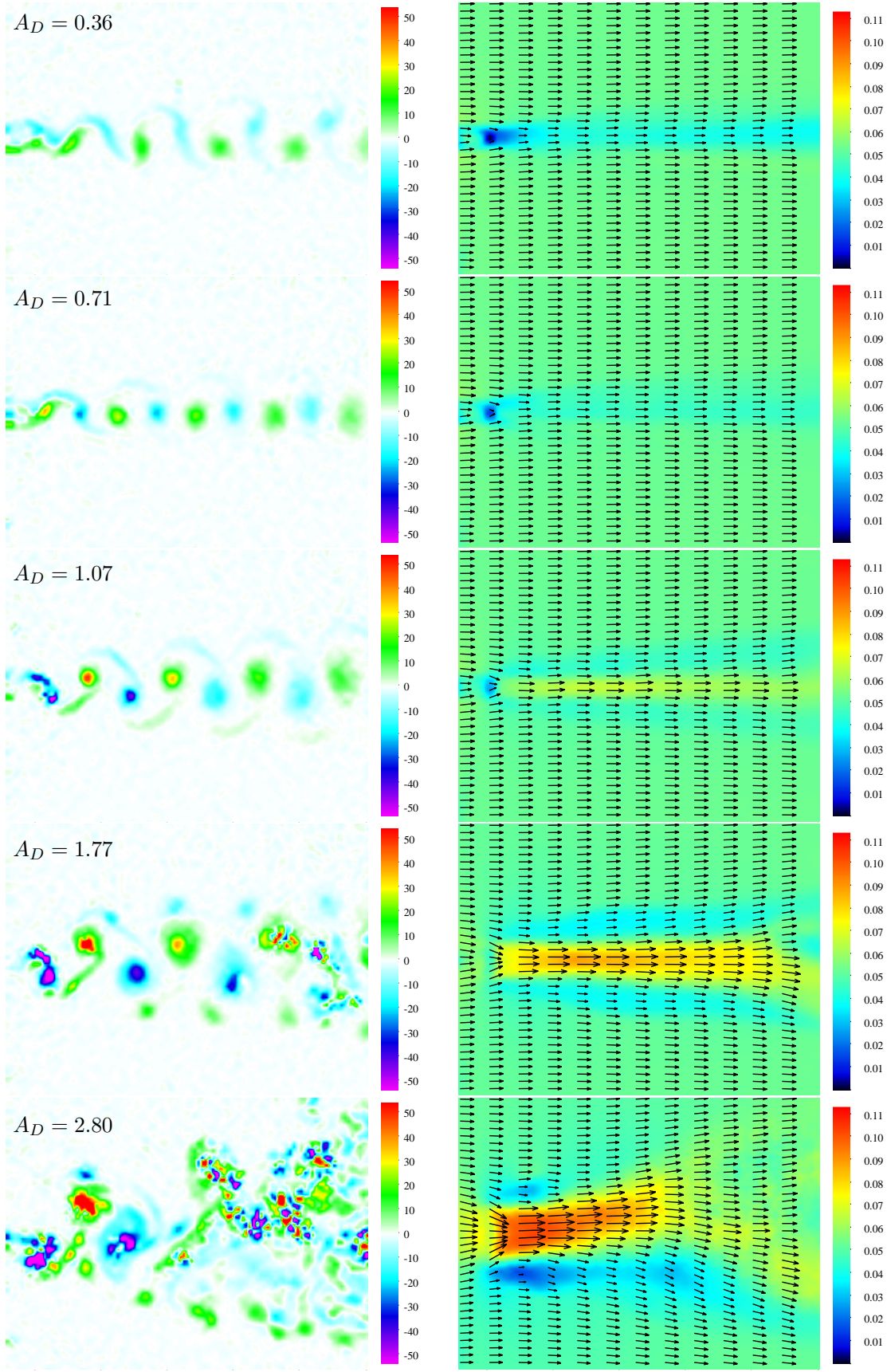}
\caption{(Color) Instantaneous spanwise vorticity fields (left column, $\mathrm{s}^{-1}$) and mean flow (time averaged horizontal velocity, right column, $\mathrm{m\; s}^{-1}$) for fixed Strouhal and Reynolds numbers ($\mathrm{Sr}=0.22$ and $\mathrm{Re}=255$) and from top to bottom, for $A_D=0.36, 0.71, 1.07, 1.77,$ and 2.8. The field of view (placed at mid-height of the foil) covers from -2$D$ to 20$D$ on the horizontal (streamwise direction $x$) and -8$D$ to 8$D$ in the vertical (crossstream direction $y$) where the origin is defined at the trailing edge of the flap at zero angle of attack.}
\label{effet_amplitude}
\end{figure*}

\section{The wake of the flapping foil}

For a fixed Reynolds number, a point in the $(\mathrm{Sr}, A_D)$ parameter space defines thus the flapping configuration and corresponds to a given vorticity signature in the wake of the foil. A parametric study as a function of the Strouhal number and the non-dimensional flapping amplitude allows to characterize the different regimes in the aforementioned $(\mathrm{Sr},A_D)$ phase space and to pin down the transition zones.

\subsection{Symmetry properties}

In Fig. \ref{effet_amplitude}, we show snapshots of spanwise vorticity (left column) as well as average horizontal velocity fields (right column), where the flapping amplitude $A_D$ is varied for a fixed value of the Strouhal number. The first row  ($A_D=0.36$) is the typical case of low-amplitude flapping which produces a forced wake with features resembling a natural B\'enard-von K\'arm\'an (BvK) vortex street, but with shedding frequency locked-in to the flapping frequency \cite[see also][]{vial2004,thiria2006,thiria2007}. In particular, the vorticity engendered at the boundary layers on each side of the foil is rolled up in vortices that, in their downstream evolution, stay on the same side of the symmetry line of the wake, defined by the axis of the foil when parallel to the free stream velocity. Accordingly, the mean flow in this case has a typical wake profile with slower speed behind the obstacle. Increasing the flapping amplitude allows us to get a very clear view of the mechanism of reversal of vortex position: for amplitude $A_D=0.71$ vortices of alternating signs align on the symmetry line of the wake (a 2S-type wake, using the nomenclature of \cite{williamson1988}), whereas for $A_D=1.07$, the vortices shed on one side of the flap organize themselves in the wake on the other side of the symmetry line constituting the reverse BvK vortex street. The mean flow picture for this transition shows that the velocity deficit in the wake profile is almost completely erased when vortices are aligned in the symmetry line ($A_D=0.71$) prior to the apparition (at $A_D=1.07$ and more clearly at $A_D=1.77$) of the jet profile that is typically associated to the reverse BvK street \cite[see also recent numerical simulations by][]{dong2006}.

Further increase in the flapping amplitude triggers a symmetry breaking in the reverse BvK wake (a feature previously observed for foils performing heaving motion \cite[][]{jones1998,lai1999,lewin2003}. %\footnote{Heaving or plunging motion refer to a lateral oscillation of the foil, perpendicular to the freestream velocity, as opposed to the present case where the oscillation is purely rotary (pitching motion).}).
A strong dipolar structure that propagates obliquely to one side of the symmetry line of the wake is formed in each flapping cycle, while a much weaker single vortex is shed on the other side. In these asymmetric wakes, the quasi-two-dimensional nature of the flow is rapidly lost and the coherence of vortical structures in the horizontal mid-plane where the observations are performed does not prevail for more than two cycles. However, the symmetry features of these wakes do seem to be defined by a quasi-two-dimensional mechanism that depends on the initial conditions: the first dipole that is formed entrains fluid behind it, deflecting the mean flow in the wake and forcing the subsequent dipoles to follow the same path. This mechanism of dipole formation resulting in a deflected wake has also been observed in soap film experiments studying the wake of oscillating cylinders \cite[][]{couder1986}, supporting the idea of a mainly 2D phenomenon. In the mean flow picture, the asymmetry of the wake can be clearly seen as a deflection of the peak in the jet profile.

\begin{figure}
\centering
\includegraphics[width=1\linewidth]{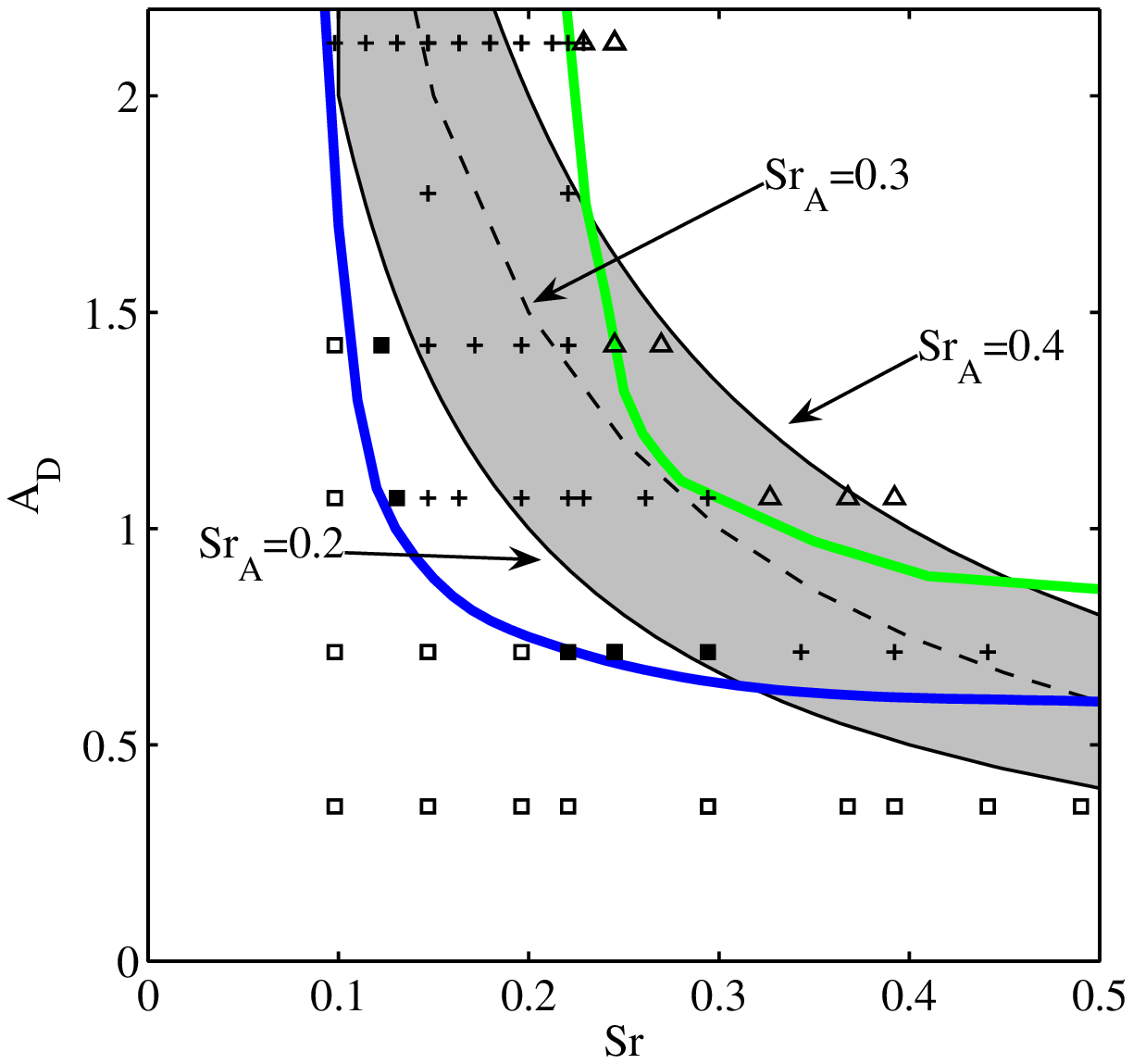}
\caption{(Color) $A_D$ vs. $\mathrm{Sr}$ map for $\mathrm{Re}=255$. Experimental points are labeled as $\square$: BvK wake; $\blacksquare$: aligned vortices (2S wake); $+$: reverse BvK wake; $\vartriangle$: deflected reverse BvK street resulting in an asymmetric wake. Blue line: transition between BvK and reverse BvK. Green line: transition between reverse BvK and the asymmetric regime. The shaded area corresponds to the $\mathrm{Sr}_A=0.3\pm 0.1$ interval, where $\mathrm{Sr}_A$ is the amplitude based Strouhal number $\mathrm{Sr}_A=\mathrm{Sr}\times A_D$ (see text).}
\label{st_vs_A_D}
\end{figure}

\subsection{The $(\mathrm{Sr},A_D)$ phase space}

The experimental phase diagram is shown in Fig. \ref{st_vs_A_D}, where different symbols represent different types of wake. Experiments where a BvK-type wake was observed ($\square$ symbols) occupy primarily the lower left region of the plot but extend to higher Strouhal numbers for the lowest amplitude tested. The transition from the BvK regime to the reverse BvK regime (blue line in Fig. \ref{st_vs_A_D}) can be univocally defined in the $(\mathrm{Sr},A_D)$ space when the position of vorticity maxima, at any given time and horizontal position, cross the symmetry axis of the wake. We remark that the transition line tends to an asymptotic value of $A_D\approx 0.6$ for $\mathrm{Sr}>0.4$ so that a threshold amplitude value exists for the production of a reverse street.
The region where the reverse BvK regime is observed ($+$ symbols) is bounded on the other side by the transition to asymmetric regimes ($\vartriangle$ symbols) represented by the green line in Fig.  \ref{st_vs_A_D}.

\subsection{The drag--thrust transition}

\begin{figure}
\centering
\includegraphics[width=1\linewidth]{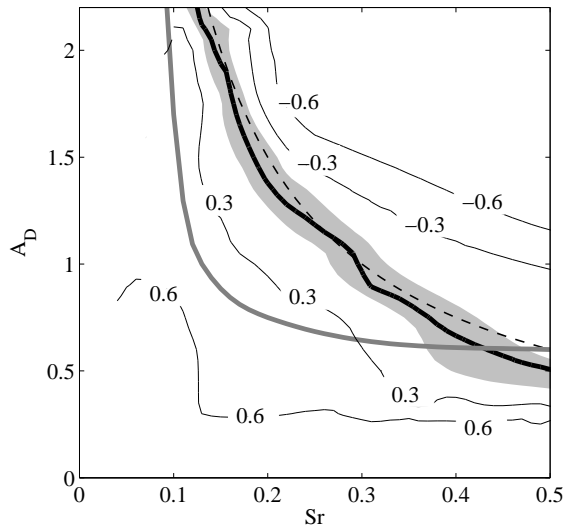}
\caption{Contours of a mean drag coefficient $C_D/C_{D0}$ surface estimated using a momentum-balance approach (see text). The black line corresponds to $C_D=0$ where the estimated drag-thrust transition occurs. The shaded area represents the estimated error for the $C_D=0$ curve due to sensitivity on the choice of the control volume. The gray line is the transition from BvK to reverse BvK (blue line in Fig. \ref{st_vs_A_D}). The dashed line corresponds to $\mathrm{Sr}_A=0.3$.}
\label{drag_coef}
\end{figure}

Using the mean velocity fields to perform a standard momentum balance in a control volume enclosing the foil (see e.g. \cite{batchelor1967}, pp. 349-351) we can obtain an indirect estimate of the mean drag. It is given by

\begin{equation}\label{drag}
    F_D=\rho U_0\int[U_0-u(y)]dy \;,
\end{equation}

\noindent where $\rho$ is the water density, $U_0$ is the free stream velocity at the center of the tunnel upstream of the flap and $u(y)$ is a velocity profile measured in the wake. The position of this wake profile is set to $X\approx 12D$ downstream from the trailing edge, where the wake can still be reasonably considered quasi-2D and across-wake pressure fluctuations can be neglected. A drag coefficient surface $C_D/C_{D0}$ obtained by interpolation of the measured points in the $(\mathrm{Sr},A_D)$ space are plotted as labeled iso-contours in Fig. \ref{drag_coef}. The drag coefficient

\begin{equation}\label{CD}
C_D=\frac{F_D}{\frac{1}{2}\rho U_0^2D}\;,
\end{equation}

normalized by its value for the non-flapping foil at zero angle of attack $C_{D0}$, allows to pin down the $(\mathrm{Sr},A_D)$ curve where $C_D$ changes sign, marking the transition between drag and propulsive regimes (heavy black line in Fig. \ref{drag_coef}). A comparison of this zero-drag curve with the transition from a BvK vortex street to a reverse BvK pattern in the wake (gray line) shows that, for increasing flapping amplitude or frequency, the reversal of the vortex street happens before the actual drag-thrust transition in almost all the parameter range studied here. This shift shows that a region in the $(\mathrm{Sr},A_D)$ plane exists where a reversed BvK pattern can be observed in the wake of the flapping foil, but the relative thrust engendered by the flapping motion is not yet enough to overcome the total mean drag. For $\mathrm{Sr}>0.4$, the two lines cross and the estimate of the drag-thrust transition stays actually slightly below the observed BvK-reverse BvK transition threshold. A possible explanation for this is that the quasi-2D hypothesis underlying the drag coefficient calculation degrades faster the higher the Strouhal number, so that 3D effects might be cooperating to
produce this result that is unexpected from a 2D point of view.

It should be noted that the approximations underlying the momentum balance used to obtain Eq. (\ref{drag}) make it a crude estimate for the mean drag on the flapping foil. In particular, as soon as the flow ceases to be parallel the momentum balance changes and Eq. (\ref{drag}) no longer holds. The mean drag in the regimes where the average wake is asymmetric and a mean lift force is likely to be non-negligible, will thus be poorly estimated by Eq. (\ref{drag}). We nonetheless verify \emph{a posteriori} that the expected drag-thrust transition does not overlap the region of the parameter space where asymmetric wakes develop, so that the prediction should be reasonable. The sensitivity of the drag coefficient calculation to the $x$-position of the wake profile, which defines the downstream boundary of the control volume, was tested performing the momentum balance at different horizontal stations in the wake. Thus, the gray `error band' around the $C_D=0$ transition line in figure \ref{drag_coef} represents the fluctuation of the transition line when the calculation of $C_D$ is performed using a wake profile measured at $X\pm4D$ (i.e. representing an overall span in the $x$ direction of $\approx 40\%$ of the length of the PIV measurement window)\footnote{A discussion of the sensitivity of the $C_D$ calculation to the choice of control volume is available as supplementary material.}.

\section{Conclusions}

The experimental study of the wake vortices produced by a high-aspect-ratio pitching foil presented in this paper gives evidence characterizing two key dynamical features relevant to wake vortex systems engendered by flapping motion: first, the transition from the well-known B\'enard-von K\'arm\'an (BvK) wake to the reversed vortex street that signals propulsive wakes, and second, the symmetry breaking of this reverse BvK pattern giving rise to an asymmetric wake. Albeit the inherently-three-dimensional nature of the wake dynamics resulting from the finite span of the foil in the hydrodynamic tunnel, the quasi-two-dimensional picture studied here by focusing on the near wake captures the essential of the vorticity dynamics because the main vortex shedding is induced when the foil motion changes direction \cite[see also][where 2D theory and simulations show good agreement with experimental studies on flapping wings]{minotti2002,wang2004}.

In conclusion, we can come back to the flapping-based propulsive regimes encountered in nature, where fish-like swimming and flapping flight occur for $\mathrm{Sr}_A=0.3\pm 0.1$. In terms of the parameters used in this paper,  $\mathrm{Sr}_A=\mathrm{Sr}\times A_D$, so that the previous interval represents a region bounded by hyperboles in the $(\mathrm{Sr},A_D)$ phase space (shaded area in Fig. \ref{st_vs_A_D}). Our first remark is that the drag-thrust transition shown in Fig. \ref{drag_coef} compares well with the $\mathrm{Sr}_A=0.3$ curve (dashed line) that characterizes animal propulsion. Within the limits imposed by the quasi-2D picture analyzed in the present work, this suggests that the conclusions obtained here should be relevant as interpretation tools in the study of natural systems. In addition, Fig. \ref{st_vs_A_D} also shows that the region defined by $\mathrm{Sr}_A=0.3\pm 0.1$ overlaps not only with the reverse BvK regime, but also with the asymmetric regime. We may conjecture that animals using flapping-based propulsion must either exploit the creation of asymmetric wakes as part of their manoeuvering techniques or, when cruising, avoid flapping regimes where the symmetry breaking of the reverse BvK street will occur.

\begin{acknowledgments}

We thank warmly D. Pradal for his invaluable help in the design and construction of the experimental setup.

\end{acknowledgments}

%\listoffigures

%\bibliographystyle{jfm}
%\bibliography{biblio_flap}

\end{document}